\documentclass[prc,aps,preprint,showkeys,lengthcheck,twocolumn,nofootinbib,notitlepage,floatfix,superscriptaddress, nolongbibliography,amsmath,amssymb]{revtex4-2}

\usepackage{graphicx}
\usepackage{dcolumn}
\usepackage{bm}
\usepackage{comment}
\usepackage[T1]{fontenc}

\usepackage{xcolor}
\usepackage{amsmath}
\usepackage{mathtools}
\usepackage{bm}
\usepackage{xcolor}

\begin{document}

\preprint{APS/123-QED}

\title{
Charmonium-nucleon femtoscopy \\as a possible probe of the nucleon gravitational form factor
}

\author{Ren~Ejima}
 \email{ejima@cns.s.u-tokyo.ac.jp}
\affiliation{%
 Center for Nuclear Study (CNS),
 University of Tokyo, 7-3-1, Hongo, Bunkyo-ku, Tokyo, Japan
 }%

\author{Daisuke~Fujii}
 \email{daisuke@rcnp.osaka-u.ac.jp}
\affiliation{%
 Advanced Science Research Center, Japan Atomic Energy Agency (JAEA),
 2-4 Shirakata Shirane, Tokai-mura, Naka-gun, Ibaraki 319-1195, Japan
}%
\affiliation{%
Research Center for Nuclear Physics (RCNP), Osaka University, Ibaraki 567-0048, Japan
}%

\author{Mamiya~Kawaguchi}
 \email{mamiya@aust.edu.cn}
\affiliation{%
 Center for Fundamental Physics, School of Mechanics and Physics,
Anhui University of Science and Technology, Huainan, 232001, People’s Republic of China
}%

\date{\today}

\begin{abstract}
We investigate the charmonium-nucleon interaction, focusing on its connection with the internal structure of the nucleon encoded in the gravitational form factors. To describe this interaction, we employ an effective potential based on the QCD multipole expansion within the leading chromoelectric dipole approximation. In this framework, the potential is expressed in terms of the energy and pressure distributions inside the nucleon. We first construct these distributions from the gravitational form factors fitted to lattice-QCD data. The remaining model parameters are then fixed by requiring that the resulting $J/\psi$-$N$ potential reproduce the HAL QCD potential outside the short-distance region, as well as the scattering phase shift estimated from the HAL QCD data. Based on this potential, we evaluate the $J/\psi$-$N$ correlation function and further investigate the $\psi(2S)$-$N$ system. We then examine the sensitivity of these correlation functions to the nucleon $D$-form factor.
\end{abstract}

\maketitle


\section{\label{intro}Introduction}
Confinement mechanism is one of the central open problems in QCD. Since quarks and gluons are confined inside hadrons and do not exist in isolation, understanding the internal structure of hadrons is essential for clarifying the dynamics of confinement. In this direction, ongoing and future experimental programs, such as those at JLab and the EIC, are expected to provide detailed information on hadronic structure through the gravitational form factors (GFFs).

The GFFs are defined as hadron matrix elements of the energy-momentum tensor (EMT) and provide access to the spatial distributions of energy, spin, pressure, and shear forces inside hadrons. In particular, the nucleon $D$-form factor, which corresponds to the spatial component of the matrix element of the EMT, characterizes the internal stress distribution, including the pressure and shear-force distributions, and is closely related to the mechanical stability of the nucleon~\cite{doi:10.1142/S0217751X18300259, RevModPhys.95.041002}. Although the forward limit of the nucleon GFFs associated with the mass and spin of the proton have been extensively studied, the $D$-term, defined at the forward limit of the $D$-form factor, has not yet been experimentally established. It is therefore often referred to as the ``last unknown global property'' of the proton~\cite{doi:10.1142/S0217751X18300259}. 

The nucleon $D$-form factor and pressure distributions have been extensively investigated using lattice QCD and various effective approaches~\cite{Ji:1997gm, Polyakov:1999gs, Belitsky:2002jp, Schweitzer:2002nm, Polyakov:2002yz, Wakamatsu:2007uc, Goeke:2007fp, Cebulla:2007ei, LHPC:2007blg, Polyakov2018, PhysRevLett.122.072003, PhysRevD.99.094026, AVELINO2019627, 10.1093/ptep/ptz093, Lorce2019, PhysRevD.101.086003, PhysRevD.101.034013, PhysRevD.102.014047, PhysRevD.102.113011, Azizi2020, PhysRevD.103.094023, PhysRevD.104.014008, PhysRevD.104.014024, PhysRevD.104.074019, OWA2022137136, PhysRevD.105.054509, PhysRevD.105.056017, 10.1093/ptep/ptac110, PhysRevD.106.076004, PhysRevD.106.114009, PhysRevD.107.054007, PhysRevD.108.094018, PhysRevD.110.L091501, PhysRevLett.132.251904, PhysRevD.109.L051502, PhysRevD.110.054005, Won2024, gzj5-7bln, FUJII2025139559, Fujii_2025, jkhv-6949, broniowski2025gravitationalformfactorsmechanical, Sugimoto_2025, nair2025protongravitationalstructuremass, Cao_2025, Dehghan2025, Yao2025, Cao20252, Kawaguchi:2025cuf, stegeman2026gravitationaldformfactor, Stegeman:2025tdl, Fukushima:2025jah, Fukushima:2026wwc}.
However, substantial theoretical uncertainties still remain, and different approaches give quantitatively different predictions for the nucleon internal properties. 
To reduce these uncertainties, experimental information on the nucleon mechanical properties is essential.

In the JLab-CLAS experiment, the quark contribution to the pressure distribution inside the nucleon was first extracted from a scattering process known as deeply virtual Compton scattering (DVCS)~\cite{Burkert2018}. Similarly, shear stresses within the nucleon have also been constrained though analyses of processes such as DVCS and near-threshold quarkonium photoproduction~\cite{burkert2021determinationshearforcesinside, Duran2023, lqv1-kf2n}.
However, it should be noted that the GFFs and the corresponding pressure distributions are not measured directly in these experiments. Rather, they are inferred from experimental observables through analyses based on generalized parton distributions (GPDs).
Since such extractions of GFFs and pressure distributions require theoretical analyses or model assumptions, some information on the underlying mechanical structure may be lost or obscured in this procedure~\cite{10.1063/1.3647111, Kumericki2016, Dupre2017, Moutarde2018, PhysRevLett.130.211902}. This motivates the search for complementary observables that can access the nucleon pressure distribution and the $D$-form factor from a different perspective.

In the light of these considerations, we explore the possibility of probing the nucleon GFFs through physical processes different from DVCS and near-threshold quarkonium photoproduction. 
In particular, we focus on the quarkonium-nucleon system where the potential can be related to the nucleon GFFs as discussed later (Section~\ref{multipoleExpansion}). Therefore this system may serve as an alternative probe for the nucleon GFFs.
To access the quarkonium-nucleon interaction, we consider femtoscopy, which is typically used to access information on hadron-hadron interactions. In recent years, led in particular by the ALICE experiment at the LHC, substantial progress has been made in measuring two-particle correlation functions, and a wide range of femtoscopic data has become available for various hadronic systems~\cite{Mikhaylov:2020hpq, Rzesa:2020iyc, Acharya_2021, ALICE:2021szj, Mihaylov:2021fmm, Humanic:2022hpq, ALICE:2021ovd, ALICE:2023wjz, ALICE:2023gxp, ALICE:2022mxo, ALICE:2022uso, Rzesa:2024dru, ALICE:2023eyl, alicecollaboration2025directaccessrho0pinteraction, ALICE:2025wuy, ALICE:2025aur, ALICE:2026vpi, ALICE:2026wwp, ALICE:2025plu, ALICE:2025kma}. In this paper, we investigate whether quarkonium-nucleon femtoscopy can serve as an alternative probe of the nucleon GFFs.

The quarkonium-nucleon system can be effectively described by soft gluonic interactions within the framework of the QCD multipole expansion. An important feature of this framework is that the resulting potential can be related to the nucleon GFFs through the energy and pressure distributions inside the nucleon. Previous studies based on this approach have investigated the $J/\psi$-$N$ correlation function~\cite{Krein2020}. However, its connection to the nucleon GFFs, especially to the $D$-form factor, has not been fully explored. In this study, we investigate the sensitivity of charmonium-nucleon correlation functions to the nucleon $D$-form factor, focusing on the $J/\psi$-$N$ and $\psi(2S)$-$N$ systems. More specifically, as a first step toward evaluating the charmonium-nucleon correlation functions, we construct the nucleon GFFs using an ansatz fitted to the lattice-QCD results~\cite{PhysRevLett.132.251904}. Using these GFFs, we obtain the energy and pressure distributions inside the nucleon and then describe the charmonium-nucleon potential. The remaining model parameters are fixed so that the resulting $J/\psi$-$N$ potential and scattering phase shift are consistent with the estimates obtained using the HAL QCD data for the $J/\psi$-$N$ potential~\cite{LYU2025139178}. With this setup, we study the sensitivity of the $J/\psi$-$N$ and $\psi(2S)$-$N$ correlation functions to the nucleon $D$-form factor.

It should be noted that the quarkonium-nucleon femtoscopy approach considered in this study also involves theoretical modeling, such as the parametrization of the nucleon GFFs. Therefore, the purpose of this study is not to extract the nucleon GFFs in a model-independent manner, but rather to examine whether the quarkonium-nucleon correlation function can be sensitive to them. Since this femtoscopic approach uses a physical process different from DVCS and near-threshold quarkonium photoproduction, it may provide a complementary perspective on how nucleon GFFs can be probed.

The remainder of this paper is organized as follows.
In Section~\ref{GFFs}, we introduce an ansatz for the nucleon GFFs and determine the parameters by fitting the ansatz to the lattice-QCD results.
In Section~\ref{potential}, we review the effective approach to the charmonium-nucleon system based on the QCD multipole expansion. We then evaluate the $J/\psi$-$N$ potential and the corresponding scattering phase shift, and discuss the applicability of the present effective-potential approach.
In Section~\ref{result}, we evaluate the charmonium-nucleon correlation functions and discuss their sensitivity to the nucleon $D$-form factor.
Finally, we summarize our findings in Section~\ref{summary}.

\section{\label{GFFs}Ansatz for the Nucleon Gravitational Form Factors and Energy and Pressure Distributions}

In this study, we analyze the charmonium-nucleon correlation function by using a gluonic effective potential derived from the QCD multipole expansion. In this framework, the effective potential is expressed in terms of the energy and pressure distributions inside the nucleon. Since these distributions are defined through the nucleon GFFs, in this section we start with the definition of the nucleon GFFs and briefly introduce the corresponding energy and pressure distributions.

The nucleon GFFs are defined through the nucleon matrix element of the EMT as
\begin{align}
    \langle N(p',s')|&T_{\mu\nu}(x)|N(p,s)\rangle\notag\\
    =&\bar{u}(p',s')\left[A(t)\frac{P_\mu P_{\nu}}{M_N}+J(t)\frac{i(P_\mu\sigma_{\nu\rho}+P_\nu\sigma_{\mu\rho})\Delta^{\rho}}{2M_N}\right.\notag\\
    &\left.+D(t)\frac{\Delta_\mu\Delta_\nu-g_{\mu\nu}\Delta^2}{4M_N}\right]u(p,s)e^{i(p'-p)x},
\end{align}
where $A(t)$, $J(t)$ and $D(t)$ are the GFFs; $\sigma_{\mu\nu}=\frac{i}{2}[\gamma^\mu,\gamma^\nu]$; $P^\mu=(p^\mu+p'^\mu)/2$ is the average momentum, $\Delta^\mu=p'^\mu-p^\mu$ is the momentum transfer, and $t=\Delta^2$; $M_N$ is a nucleon mass; and $u(p,s)$ is the Dirac spinor with spin $s$. In the following, we focus on the case with $s=s'$.

To parametrize the momentum-transfer dependence of the GFFs, we assume the ansatz\footnote{Perturbative-QCD counting rules predict, up to logarithmic corrections, $A(t), \ J(t)\sim 1/t^2, \ D(t)\sim 1/t^3$ at asymptotically large $|t|$ \cite{Tong:2022zax}. For simplicity, we employ tripole parametrizations for all GFFs, which therefore do not reproduce the perturbative asymptotic behavior of $A(t)$ and $J(t)$. We have also performed the analysis using dipole parametrizations for $A(t)$ and $J(t)$, consistent with the perturbative power behavior, and confirmed that the resulting changes are mainly confined to the short-distance region and do not qualitatively affect our main conclusion concerning the sensitivity of the correlation function to the $D$-term. More fundamentally, the effective charmonium--nucleon interaction employed in Sec.~III is constructed within the leading soft-gluon multipole approximation, while effects beyond this approximation become increasingly relevant in the very-short-distance region. Therefore, within the present framework, enforcing the perturbative asymptotic behavior of the input GFFs specifically to refine the potential in this region would have only limited physical significance.}
\begin{equation}
\begin{split}
\label{eq:GFFsAnsatz}
    A(t)=\frac{1}{(1-t/\Lambda_A^2)^3}\\
    J(t)=\frac{1/2}{(1-t/\Lambda_J^2)^3}\\
    D(t)=\frac{D_0}{(1-t/\Lambda_D^2)^3},
\end{split}
\end{equation}
where $\Lambda_{A,J,D}$ are model parameters and $D_0$ represents the $D$-term defined by the forward-limit value of the $D$-form factor, $D_0=D(0)$. 
We determine these parameters by fitting the parametrizations in Eq.~\eqref{eq:GFFsAnsatz} to the lattice-QCD GFF data reported in Ref.~\cite{PhysRevLett.132.251904}, yielding
$\Lambda_A=1.75$ GeV, $\Lambda_J=1.93$ GeV, $\Lambda_D=1.01$ GeV, and $D_0=-2.69$.

\begin{figure}[t!]
\begin{center}
\includegraphics[width=80mm]{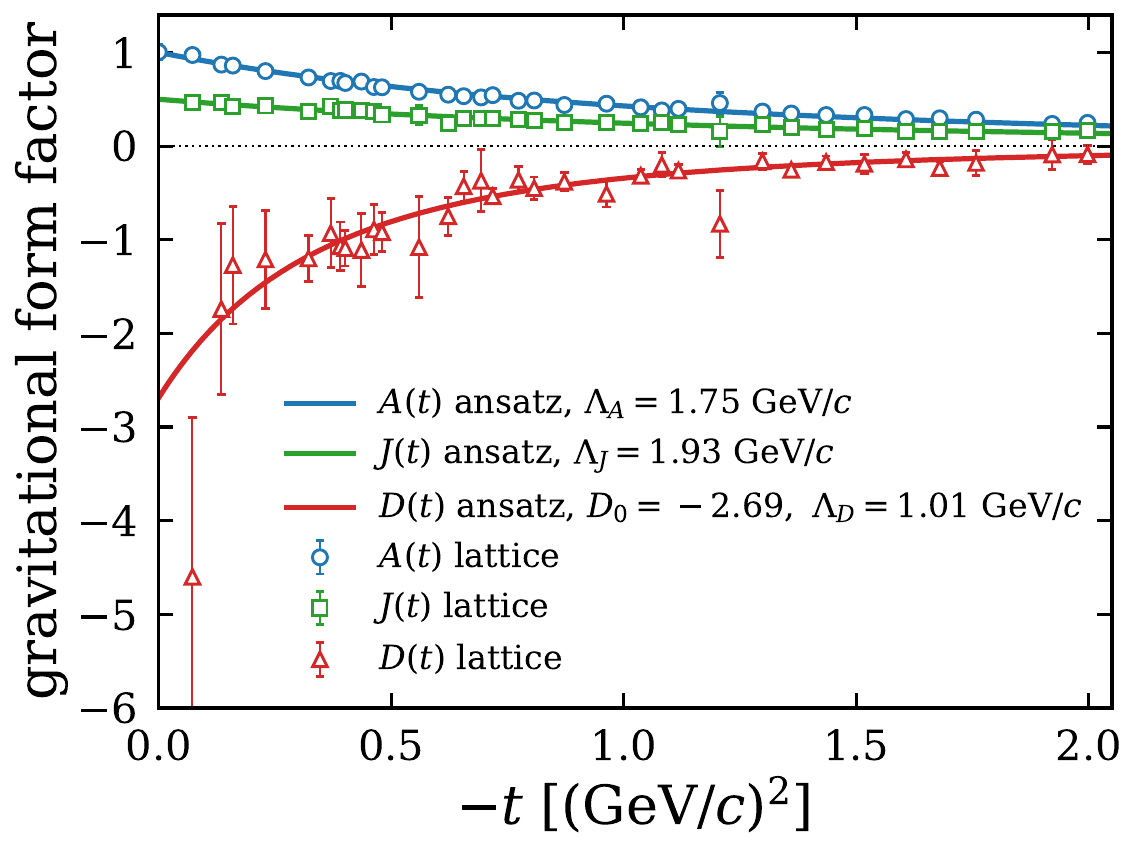}%
\caption{\label{fig:GFFs}
Momentum transfer dependence of the GFFs based on the ansatz in Eq.~\eqref{eq:GFFsAnsatz}, fitted to the lattice QCD results~\cite{PhysRevLett.132.251904}. 
}
\end{center}
\end{figure}

As seen in the Fig.~\ref{fig:GFFs}, the ansatz in Eq.~\eqref{eq:GFFsAnsatz} gives a reasonable description of the momentum-transfer dependence of the lattice-QCD results, although the $D$-form factor has a relatively large uncertainty around the forward limit.

\begin{figure}[t!]
\begin{center}
\includegraphics[width=80mm]{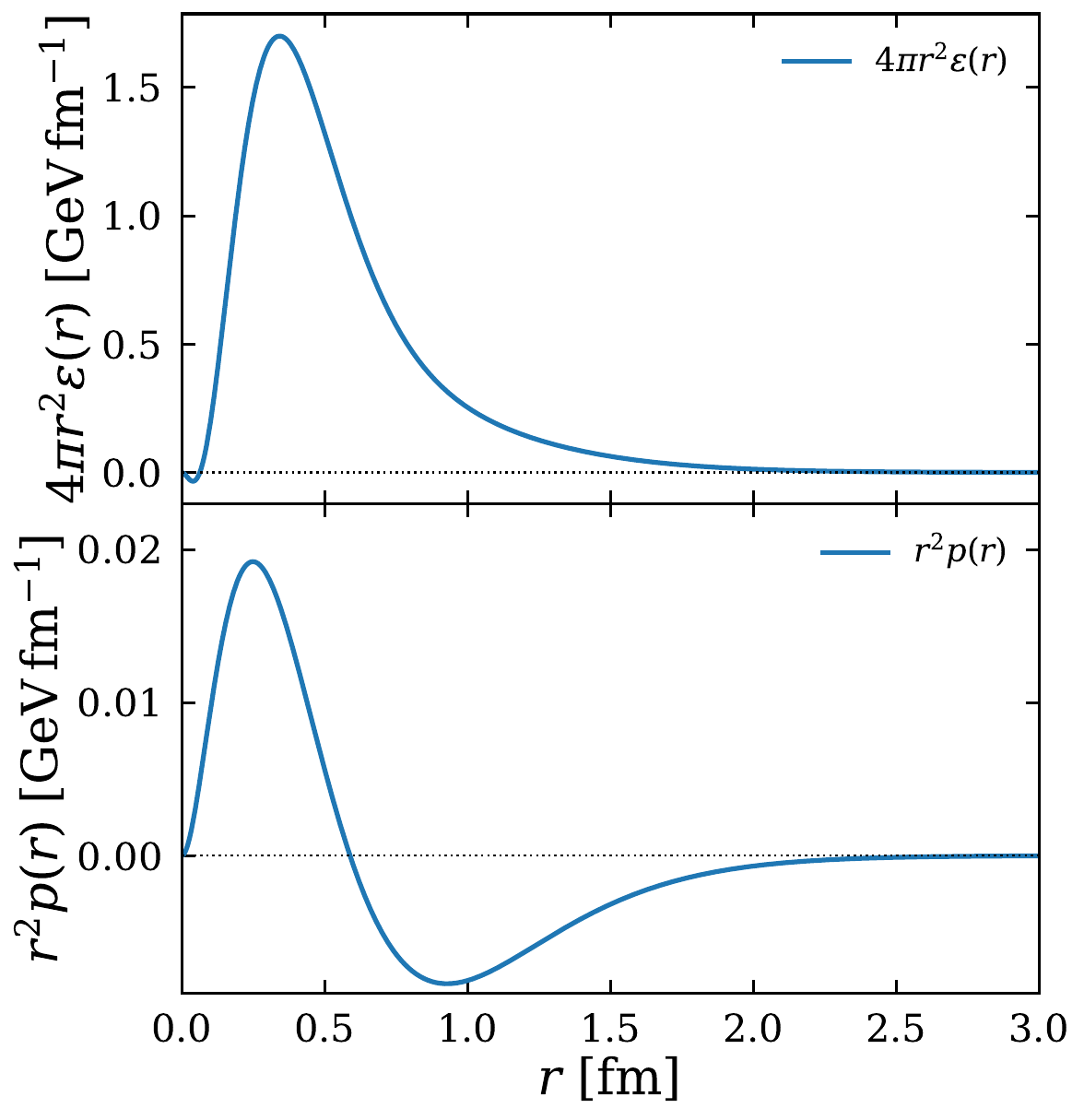}%
\caption{\label{fig:E_and_P_dist}
The radial dependence of the energy distribution (top) and pressure distribution (bottom) inside a nucleon, calculated from the ansatz in Eq.~\eqref{eq:GFFsAnsatz}.
}
\end{center}
\end{figure}

The energy and pressure distributions inside the nucleon are obtained from the Fourier transform of the matrix element of the EMT in the Breit frame, where $P^\mu=(P^0,0,0,0)$ and $\Delta^\mu=(0,\vec \Delta)$:
\begin{align}
    \tilde{T}_{\mu\nu}(\vec{r})&=\int\frac{d^3\vec\Delta}{(2\pi)^3}e^{-i\vec r\cdot\vec\Delta}\frac{\langle N(p')|T_{\mu\nu}(x=0)|N(p)\rangle}{{\bar u(p')u(p)}}.
\end{align}
Using this static EMT, the energy distribution $\epsilon(r)$ and pressure distribution $p(r)$ are introduced as
\begin{align}
\epsilon(r)&=\tilde T_{00}(r)\notag\\
p(r)&=\frac{1}{3}\delta^{ij}\tilde T_{ij}(r).
\label{eq:FT}
\end{align}
The numerical results are shown in Figs.~\ref{fig:E_and_P_dist}.
These distributions are used in the evaluation of the charmonium-nucleon potential discussed in the following section.

\section{\label{potential}
Effective framework for the charmonium-nucleon system
}
In this section, we first briefly review the effective approach to the charmonium-nucleon system based on the QCD multipole expansion. We then evaluate the $J/\psi$-$N$ potential and compare it with the lattice-QCD result to examine the applicability of the present approach. We also examine the corresponding scattering phase shift by solving the Schr\"{o}dinger equation.

\subsection{Charmonium-nucleon potential from the QCD multipole expansion}\label{multipoleExpansion}

Since the charmonium states such as $J/\psi$ and $\psi(2S)$ are compact bound states of a heavy charm quark and antiquark, their low-energy interactions with a nucleon are mediated by soft gluonic fields and are expected to be described in terms of the dipole-type interaction of the compact charmonium state with these gluonic fields. This picture can be formulated more systematically using the QCD multipole expansion, which is based on the methodology of the operator product expansion~\cite{PESKIN1979365, BHANOT1979391}. At the leading order in this framework, the low-energy nucleon-charmonium scattering amplitude is factorized into the  chromoelectric polarizability of the $nS$ charmonium state,
$\alpha_{nS}$, and the nucleon matrix element of a gluonic operator,
\begin{equation}
    \mathcal{M}=\frac{1}{2}\alpha_{nS}\langle N(p')|{\bm E}^2|N(p)\rangle,
\label{scat_amp}
\end{equation}
where ${\bm E}$ represents the chromoelectric field (the strong coupling constant is absorbed). 

The diagonal and nondiagonal chromoelectric polarizabilities of physical charmonia are still not precisely determined. In the heavy-quark and large-$N_c$ limits, they can be evaluated perturbatively by treating the quarkonium as a Coulombic system~\cite{BHANOT1979391,PhysRevD.93.054039}. The leading-order result is
\begin{align}
\alpha_{nS}=\frac{16\pi n^2}{3g^2N_c^2}c_na_0^3,
\label{eq}
\end{align}
where $c_1=7/4$, $c_2=251/8$, and $a_0=16\pi/(g^2N_cm_Q)$ is the Bohr radius of the quarkonium, with $g$ being the strong coupling constant evaluated at the scale of the quarkonium radius and the charm quark mass $m_Q$. Using a Bohr radius obtained by fitting the $J/\psi$ and $\psi(2S)$ masses within the Coulombic description, Ref.~\cite{PhysRevD.93.054039} estimated $\alpha_{1S}\simeq0.2~{\rm GeV}^{-3}$, $\alpha_{2S}\simeq12~{\rm GeV}^{-3}$.
However, these estimates are valid within the heavy-quark, large-$N_c$, and Coulombic approximations and depend strongly on the input parameters. In fact, another evaluation within a similar framework gives $\alpha_{1S}\simeq4.1~{\rm GeV}^{-3}$~\cite{FERRETTI2018702}. Therefore, the perturbative values obtained from Eq.~\eqref{eq} should not be regarded as quantitatively well-determined values for physical charmonium.

Several attempts have been made to determine the chromoelectric polarizabilities phenomenologically. Combining a quarkonium--nucleon effective field theory with lattice-QCD data gives $\alpha_{1S}\simeq0.24~{\rm GeV}^{-3}$~\cite{PhysRevD.98.014029}. An analysis of the lattice-QCD $J/\psi$--$N$ potential gives~\cite{PhysRevD.98.034030}
\begin{align}
\alpha_{1S}=1.6\pm0.8~{\rm GeV}^{-3}.
\end{align}
More reliable phenomenological information is available for the nondiagonal polarizability from the decay $\psi(2S)\to J/\psi\pi\pi$, for which $|\alpha_{2S\to1S}|=(1.18\pm0.01\pm0.05)~ \rm GeV^{-3}$ has been extracted from experimental data~\cite{Guo:2004dt,PhysRevD.74.014011,Yan:2026oil}. In contrast, no comparably reliable determination of the diagonal polarizability $\alpha_{2S}$ is currently available.

In the present study, we therefore use $\alpha_{1S}=1.6~{\rm GeV}^{-3}$, following Ref.~\cite{PhysRevD.98.034030}, which is directly related to the $J/\psi$--$N$ potential considered here. For $\alpha_{2S}$, we instead vary its value as a parameter because no sufficiently reliable determination is presently available. The Cauchy--Schwarz inequality, $\alpha_{1S}\alpha_{2S}\geq|\alpha_{2S\to1S}|^2$, may provide an additional constraint on $\alpha_{2S}$ once $\alpha_{1S}$ and $\alpha_{2S\to1S}$ are determined more accurately. A more precise and independent determination of the diagonal chromoelectric polarizabilities, from lattice QCD or other phenomenological approaches, will therefore be important for reducing the uncertainty of quarkonium--nucleon interactions.

The nucleon matrix element in Eq.~\eqref{scat_amp} can be expressed in terms of the energy and pressure distributions inside the nucleon. To see this, the chromoelectric field operator $\bm{E}^2$ is rewritten in terms of the gluon field strength $G^\alpha_{\mu\nu}$ and the gluonic part of the QCD energy-momentum tensor, $T_{00}^G$, as
\begin{align}
\bm{E}^2
    &=-\frac{1}{4}g^2G^\alpha_{\mu\nu}G^{\alpha\mu\nu}+g^2T_{00}^G\label{eq:E2}.
\end{align}
Furthermore, the first term can be related to the trace anomaly. 
In general, the trace anomaly consists of the gluonic anomalous contribution and the quark-mass contribution.
A recent study based on the skyrmion approach suggests that the nucleon matrix element of the quark-mass contribution is relatively small compared with that of the gluonic anomalous contribution~\cite{gzj5-7bln}.
Therefore, the quark-mass contribution is expected to play only a minor role in Eq.~\eqref{scat_amp}.
Motivated by this expectation, we take, for simplicity, the chiral limit in the light-quark sector, including the heavier-quark contributions. In this limit, the trace of the EMT relevant to the nucleon matrix element is given solely by the gluonic anomalous operator, $T^{\mu}\,_{\mu}=(\beta(g_s)/{2g_s})G^\alpha_{\mu\nu}G^{\alpha\mu\nu}$,
where $g_s$ is the strong coupling constant renormalized at a low-energy scale, and $\beta(g_s)$ is the beta function of $g_s$.

At the leading order in the $\beta$ function, the scattering amplitude becomes
\begin{align}
    \mathcal{M}=
    \frac{1}{2}\alpha_{nS}\left[\frac{1}{4}\cdot\frac{32\pi^2}{bg_s^2}g^2\langle N(p')|T^{\mu}\,_{\mu}|N(p)\rangle\right.\notag\\
    \left.+g^2\langle N(p')|T_{00}^G|N(p)\rangle\right],\label{eq:amp2}
\end{align}
with $b = (11N_c - 2N_f)/3$, 
$N_c = 3$ and $N_f = 3$.
In the conventional treatment, the gluonic part of the EMT, $T_{00}^G$, appearing in the second term is approximated by the full EMT $T_{00}$ with a phenomenological parameter: $\langle N'|T_{00}^{G}|N\rangle=\xi\langle N'|T_{00}|N\rangle$.
Meanwhile, recent lattice-QCD simulations~\cite{PhysRevLett.132.251904} have provided information on the nucleon matrix element of $T_{00}^G$ itself.
Thus, in principle, introducing the parameter $\xi$ is no longer necessary.
However, a direct use of the lattice data would require additional Ansätze for the gluonic GFFs, as in Eq.~\eqref{eq:GFFsAnsatz}, with several model parameters. To avoid this complication associated with a multiparameter fit of the gluonic GFFs, we adopt the conventional prescription with a single parameter $\xi$.

To discuss the nucleon-charmonium interaction,  we convert the scattering amplitude into an effective potential by taking the Fourier transform with respect to the momentum transfer $\vec \Delta$. Using Eq.~\eqref{eq:FT}, 
we arrive at the conventional nucleon-charmonium potential based on the QCD multipole expansion, 
\begin{equation}\label{eq:potential}
    V(r)=-\alpha_{nS}\frac{4\pi^2}{b}\left(\frac{g^2}{g_s^2}\right)\left[\nu\epsilon(r)
    -3p(r)\right].
\end{equation}
with $\nu=1+\xi bg_s^2/(8\pi^2)$. Note that the radial coordinate $r$ in this potential originates from the Fourier transform of the nucleon matrix element and denotes the spatial coordinate of the energy and pressure distributions of nucleon. Therefore, it is conceptually different from the relative distance $r^*$ between the charmonium and the nucleon. In the present analysis, the charmonium is treated as a pointlike particle located at a distance $r$ from the center of the nucleon, and then the radial coordinate $r$ is identified with the relative distance, $r=r^*$.

\subsection{Numerical evaluation of $J/\psi$-$N$  potential}

To perform the numerical analysis based on the nucleon-charmonium potential in Eq.~\eqref{eq:potential}, we specify the parameter values used in this framework. 
For the $J/\psi$-$N$ system, we use the estimate
$\alpha_{1S}=1.6\pm0.8,{\rm GeV}^{-3}$ as a reference value, as discussed above.
Indeed, this value was evaluated using
$g^2/g_s^2=1.37\pm0.37$ and $\nu=1.5\pm0.1$
in the previous study~\cite{PhysRevD.98.034030}. However, these parameters still involve sizable uncertainties. Therefore, in this study, we adopt 
\begin{equation}
(\alpha_{1S},\, g^2/g_s^2,\, \nu)
=
(1.6\,{\rm GeV}^{-3},\,1,\,1.5)
\label{set_parameter}
\end{equation}
as a reference parameter set.
This set is selected to reproduce the scattering phase shift obtained from the lattice-QCD data~\cite{LYU2025139178}, as discussed later.
Given the fact that the correlation function is constructed from the relative wave function of the $J/\psi$-$N$ pair (see Eq.\eqref{eq:KP}), reproducing the phase shift serves as a useful benchmark for studying the correlation function.

\begin{figure}[t!]
\begin{center}
\includegraphics[width=80mm]{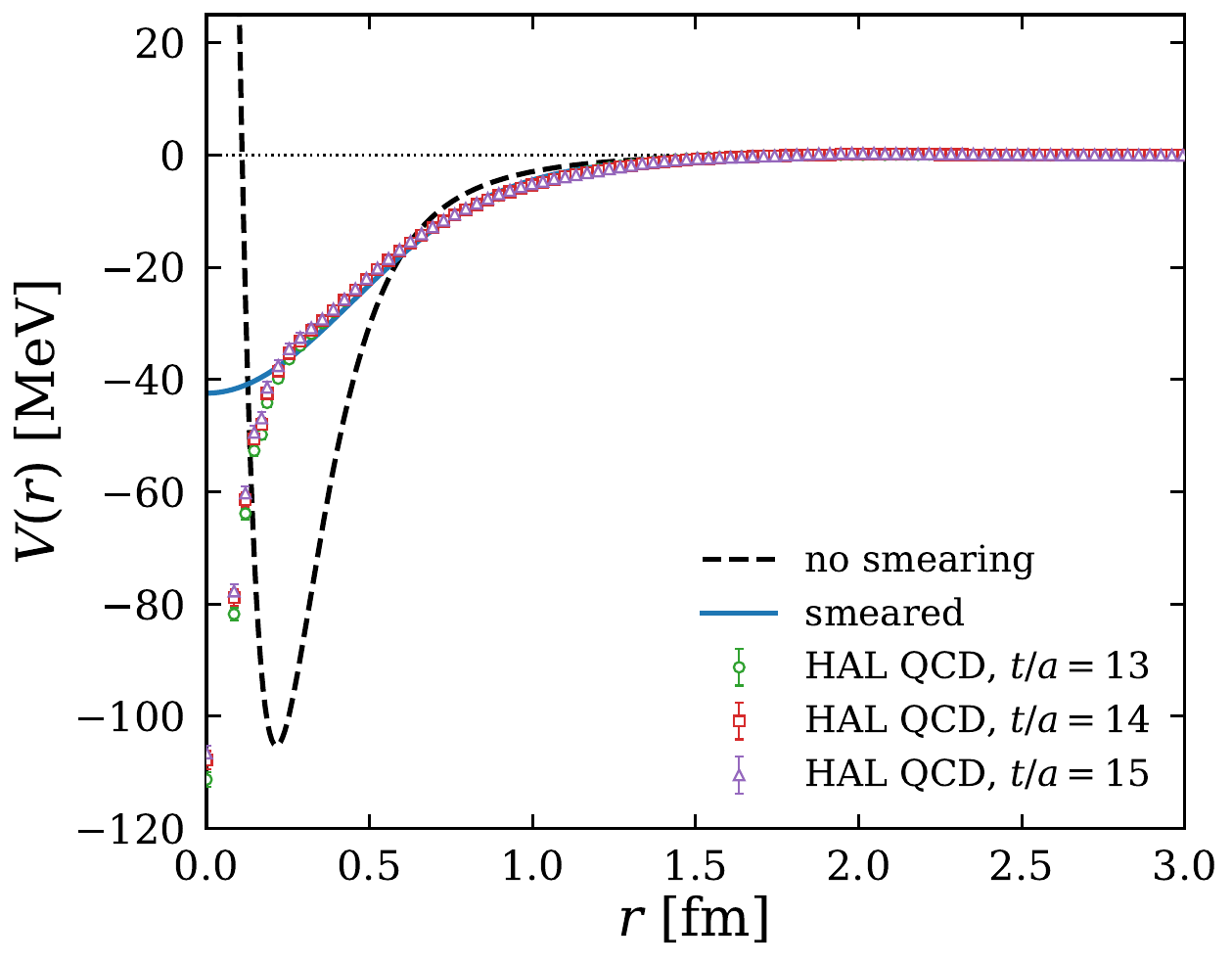}%
\caption{\label{fig:HALQCDpotential}
The $J/\psi$-$N$ potential. 
The dashed line shows the potential calculated using Eq.~\eqref{eq:potential}, while the blue solid line shows the smeared potential given in Eq.~\eqref{eq:smear}. These potentials are compared with the averaged potential evaluated from the HAL QCD data~\cite{LYU2025139178}. 
The values of $t/a$ shown for the HAL QCD data denote the ratio of Euclidean time separation and lattice spacing.
}
\end{center}
\end{figure}

Using the parameter set in Eq.~\eqref{set_parameter}, the energy-density and pressure distributions in Fig.~\ref{fig:E_and_P_dist}, 
we numerically evaluate the $J/\psi$-$N$ potential.
The resulting potential is presented in Fig.~\ref{fig:HALQCDpotential} (black dashed line) together with the lattice-QCD result~\cite{LYU2025139178}. 
In the HAL QCD study~\cite{LYU2025139178}, the potential has been calculated separately for the spin-$1/2$ and spin-$3/2$ channels. For comparison with our result, we construct the spin-averaged potential: 
$V_{\rm{spin\,avg}}=\frac{1}{3}V_{(s=1/2)}+\frac{2}{3}V_{(s=3/2)}$.
As shown in Fig.~\ref{fig:HALQCDpotential}, the observed potential is consistent with the lattice-QCD result in the long-distance region. 
However, a significant deviation appears in the short-distance region below $1\,{\rm fm}$. 

At short quarkonium--nucleon separations, finite-size, nonlocal, and higher-multipole effects beyond the leading soft-gluon multipole approximation may become relevant. To treat such short-distance effects in a simple phenomenological manner, we consider the following simple spatial distribution of quarkonium:
\begin{align}
    \rho_{Q\bar{Q}}=\frac{1}{(2\pi \sigma^{2}_{Q\bar{Q}})^{3/2}}{\rm{exp}}\left(-\frac{r^2}{2\sigma^2_{Q\bar{Q}}}\right),
\end{align}
where $\sigma_{Q\bar{Q}}$ characterizes the spatial size of the quarkonium state. In the present work, we take $\sigma_{J/\psi}=0.3\,{\rm{fm}}$ as typical values for the corresponding charmonium sizes.
By smearing the potential with this quarkonium spatial distribution, we phenomenologically incorporate the short-distance effects discussed above, in particular those associated with the finite spatial size of the quarkonium. The smeared potential is then defined as
\begin{align} 
    V_{\rm{smear}}(r)=\int d^3r'\rho_{Q\bar{Q}}(|r-r'|)V(r')\label{eq:smear}.
\end{align}
After this smearing procedure, the resulting potential becomes consistent with the lattice-QCD result down to $r\simeq0.3{\rm fm}$, as shown in Fig.~\ref{fig:HALQCDpotential} (blue solid line).

\subsection{$J/\psi$-$N$ scattering phase shift}

\begin{figure}[t!]
\begin{center}
\includegraphics[width=80mm]{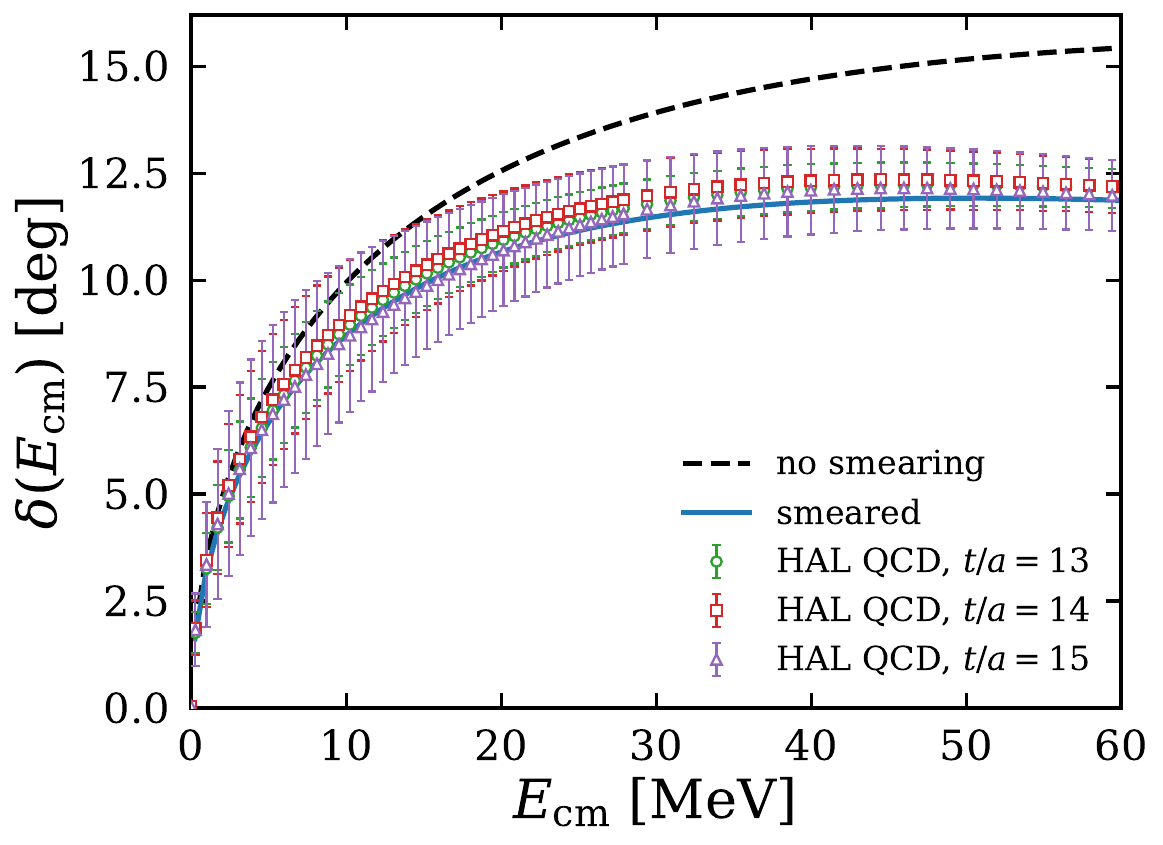}%
\caption{\label{fig:HALQCDphaseshift}
The $J/\psi$-$N$ scattering phase shifts as a function of $E_{\rm cm}$. The results obtained in this work are compared with those evaluated by solving the Schr\"{o}dinger equation with the spin-averaged HAL QCD potential shown in Fig.~\ref{fig:HALQCDpotential}.
}
\end{center}
\end{figure}

With the $J/\psi$-$N$ potential at hand, we next solve the Schr\"{o}dinger equation for the relative motion of the charmonium-nucleon two body system,
\begin{align}
    \left[-\frac{\nabla^2}{2\mu}+V(r)\right]\psi(r,k^*)=
    E_{\rm cm} \psi(r,k^*),
\end{align}
where $\mu = m_{\bar{Q}Q}M_N / (m_{\bar{Q}Q} + M_N)$ is the reduced mass of the charmonium-nucleon system, with $m_{\bar{Q}Q}$ being the charmonium mass.
The relative energy $E_{\rm cm}$ is given by $E_{\rm cm}=k^{*2}/(2\mu)$ with $k^*$ being the relative momentum. The relative wave function of the $J/\psi$-$N$ pair is written as  $\psi(r,k^*)$, reflecting its dependence on $k^*$.
Since the HAL QCD calculation suggests that the $s$-wave contribution dominates the low-energy $J/\psi$-$N$ scattering~\cite{LYU2025139178}, we focus on the $s$-wave component in what follows. The wave function is then written as
\begin{align}
 \psi(r,k^*)= \frac{1}{\sqrt{4\pi}}\frac{u(r,k^*)}{rk^*},
\end{align}
where $u(r,k^*)$ satisfies the following differential equation: $\frac{d^2u(r,k^*)}{dr^2}+[k^{*2}-2\mu V]u(r,k^*)=0$.
For the boundary condition at the origin, we impose $u(r\rightarrow0)\sim r$ to avoid a divergence at the origin. 
At large distances, the wave function is normalized such that it approaches the free spherical wave in the absence of the interaction, $\psi_{\rm{free}}(r,k^*)={\rm{sin}}(rk^*)/rk^*$. The scattering phase shift is then extracted from the asymptotic behavior of the interacting wave function. The resulting phase shift is shown in Fig.~\ref{fig:HALQCDphaseshift}.
The phase shift obtained with the bare potential deviates from the lattice-QCD result, while the smeared potential  yields the result much closer to the estimation from the HAL-QCD data~\cite{LYU2025139178}.
This indicates that the reference parameter set in Eq.~\eqref{set_parameter} is appropriate, and the corresponding wave function provides a reasonable description of the $J/\psi$-$N$ two-body system.

\section{\label{result} 
Charmonium-nucleon correlation function and its sensitivity to the nucleon $D$-term}

In this section, we evaluate the charmonium–nucleon correlation function using the wave function obtained in the previous section. The correlation function is calculated using the Koonin–Pratt formula,
\begin{equation}\label{eq:KP}
    C(k^*)=1+\int dr^3 S(r)\left(|\psi(r,k^*)|^2-|\psi_{\rm{free}}(r,k^*)|^2\right),
\end{equation}
where $S(r)$ is the source function assumed as a Gaussian source,
$S(r)=\exp\left[-r^2/(4R^2)\right]/(4\pi R^2)^{3/2}$. Here, the source size is set to $R=1.0\,{\rm{fm}}$, which is a typical value for pp collisions.


\subsection{
$J/\psi$-$N$ correlation function
}

\begin{figure}[t!]
\begin{center}
\includegraphics[width=80mm]{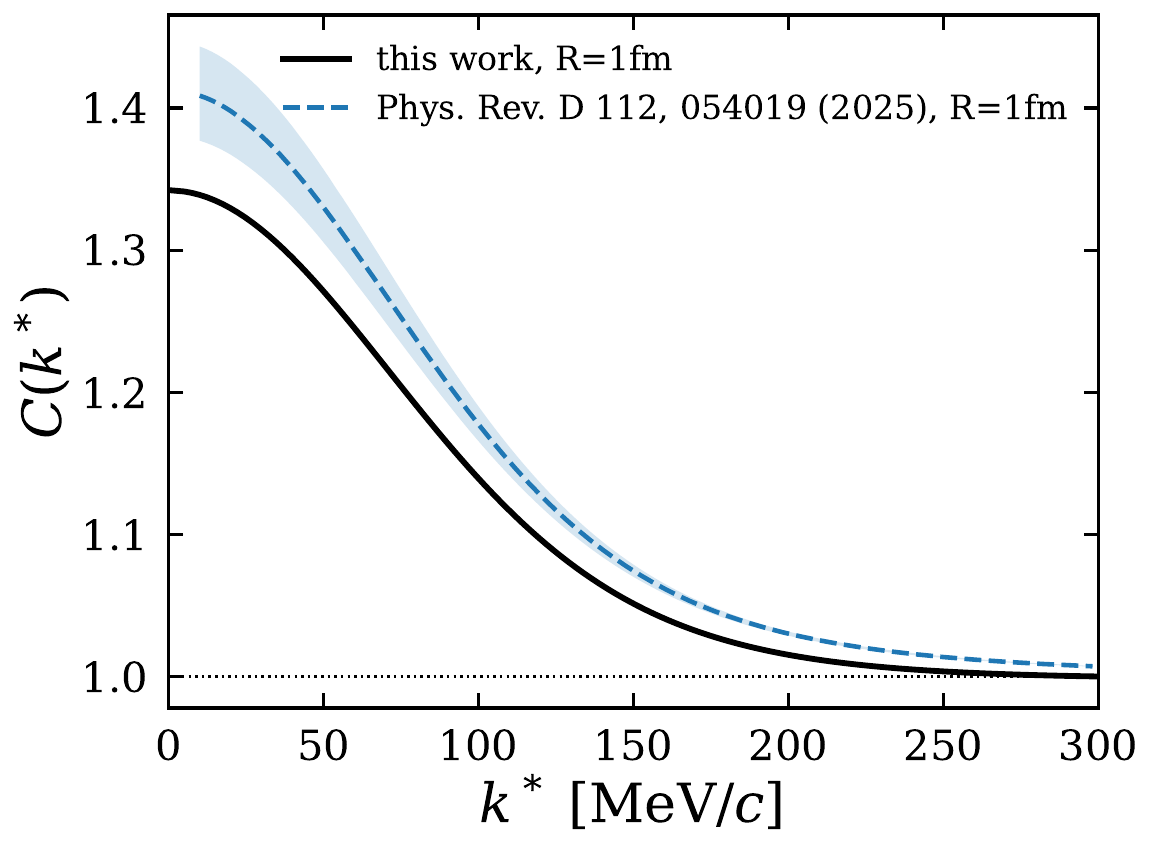}%
\caption{\label{fig:JpsiBaseline}
The $J/\psi$-$N$ correlation function as a function of $k^*$. The black solid line is obtained using the smeared $J/\psi$--$N$ potential, and the result is compared with the model-independent analysis based on the HAL QCD potential~\cite{3bdh-blwh}.
}
\end{center}
\end{figure}

We first present the $J/\psi$-$N$ correlation function calculated using the wave function obtained from the smeared potential, as shown in Fig.~\ref{fig:JpsiBaseline}. 
The resulting correlation function is larger than $C=1$, indicating the attractive $J/\psi$-$N$ interaction. 
As $k^*$ increases, the correlation function smoothly decreases and approaches $C=1$. 
This behavior is qualitatively consistent with the spin-averaged result obtained from the HAL QCD data using a model-independent analysis~\cite{3bdh-blwh}.

\begin{figure}[t!]
\begin{center}
\includegraphics[width=80mm]{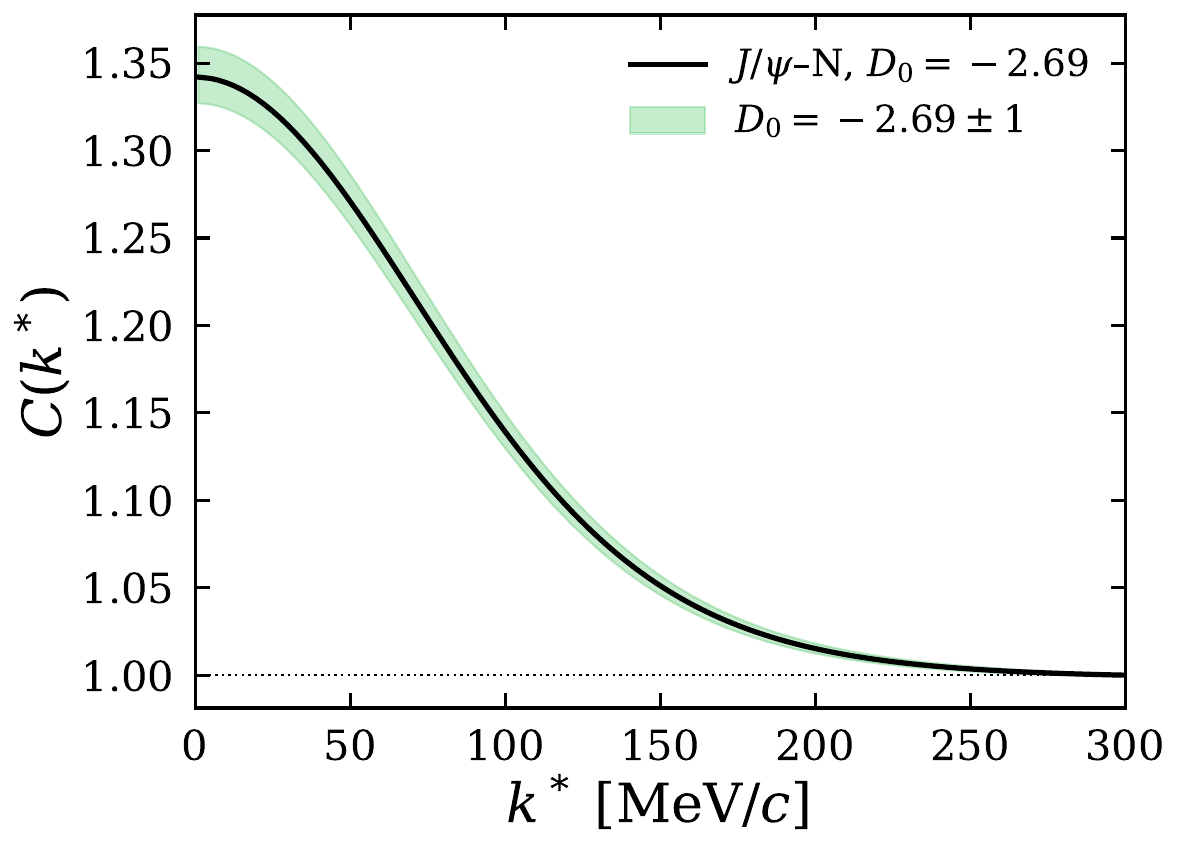}%
\caption{\label{fig:JpsiCF_Dterm}
The $J/\psi$-$N$ correlation function. The shaded band represents the variation of the correlation function when the $D$-term parameter is varied within $D_0=-2.69\pm1$.
}
\end{center}
\end{figure}

In Fig.~\ref{fig:JpsiBaseline}, we have used the value $D_0=-2.69$, which has been fixed from the lattice QCD data for the GFFs~\cite{PhysRevLett.132.251904}, as discussed in Sec.~\ref{GFFs}. 
However, the lattice results still have sizable uncertainties in the $D$-form factor near the forward limit~\cite{PhysRevLett.132.251904}.
Therefore, in the present study, we vary the value of $D_0$ and examine the sensitivity of the $J/\psi$-$N$ correlation function to the $D$-term. Specifically, we vary $D_0$ within the range $D_0=-2.69\pm1$ and the resulting correlation functions are shown in Fig.~\ref{fig:JpsiCF_Dterm}. As can be seen from the figure, the correlation function exhibits a weak dependence on $D_0$, indicating that the $J/\psi$-$N$ correlation function does not directly reflect the magnitude of $D_0$.

\subsection{$\psi(2S)$-$N$ correlation function}

Within the potential analysis based on the QCD multipole expansion, the magnitude of the charmonium-nucleon potential is strongly affected by the chromoelectric polarizability $\alpha_{nS}$, as seen from Eq.~\eqref{eq:potential}. Since the pressure distribution is related to the $D$-term through the Fourier transform of the nucleon EMT matrix element, the $D$-term dependence of the potential can be enhanced with a large $\alpha_{nS}$. Indeed, a previous study has suggested that the chromoelectric polarizability of $\psi(2S)$ can take a large value, $\alpha_{2S}=24\pm12,{\rm GeV^{-3}}$~\cite{PhysRevD.98.034030}. This motivates us to turn to the $\psi(2S)$-$N$ system, where the $D$-term dependence is expected to be more clearly reflected in the charmonium-nucleon correlation function.

\begin{figure}[t!]
\begin{center}
\includegraphics[width=80mm]{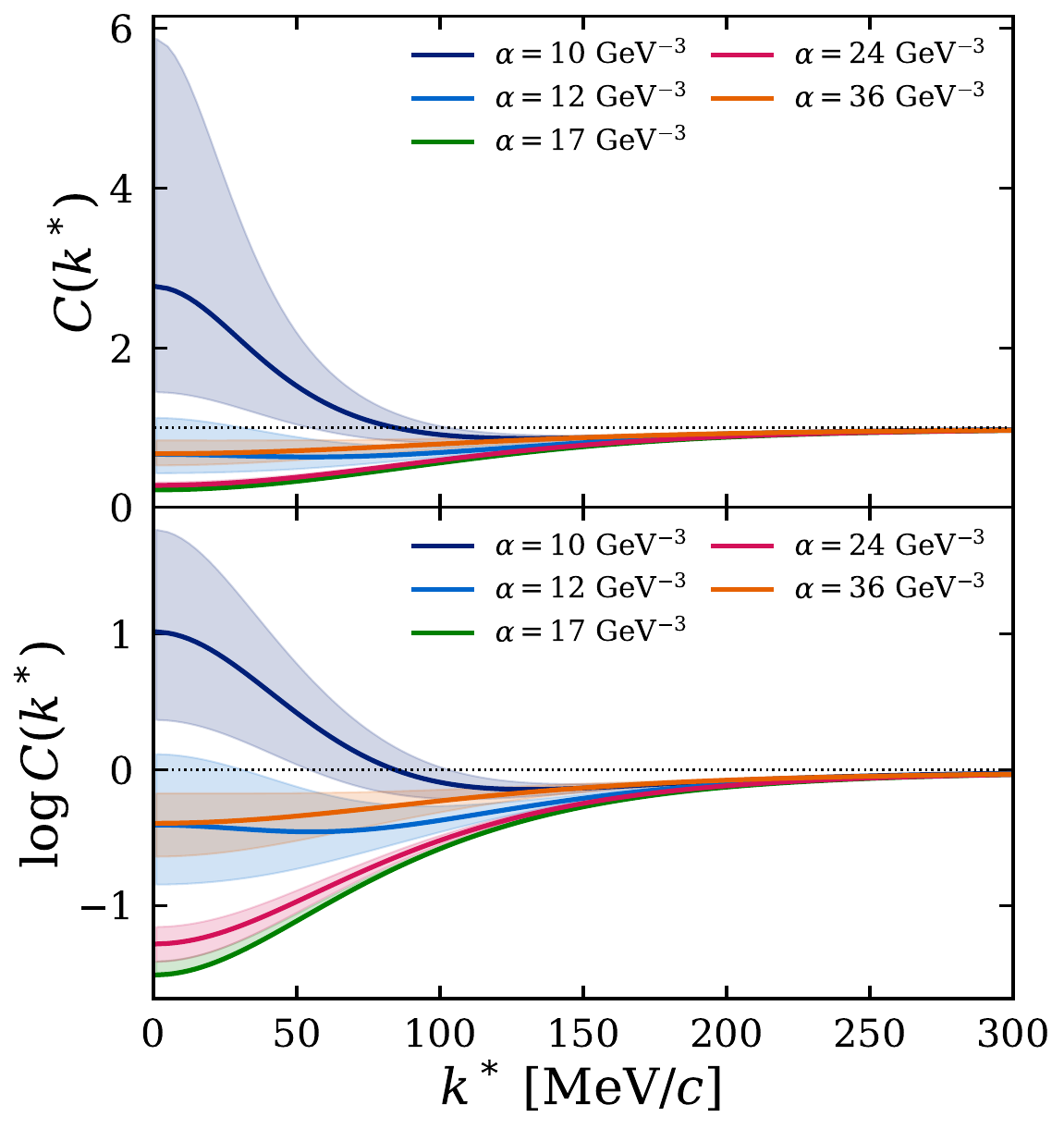}%
\caption{\label{fig:psi2SCF_Dterm}
The $\psi(2S)$-$N$ correlation function for $\alpha_{2S}$. The top and bottom panels show $C(k^*)$ and $\log C(k^*)$, respectively. The shaded bands represent the variation of the correlation function when the $D$-term parameter is varied within $D_0=-2.69\pm1$.
}
\end{center}
\end{figure}

\begin{figure}[t!]
\begin{center}
\includegraphics[width=80mm]{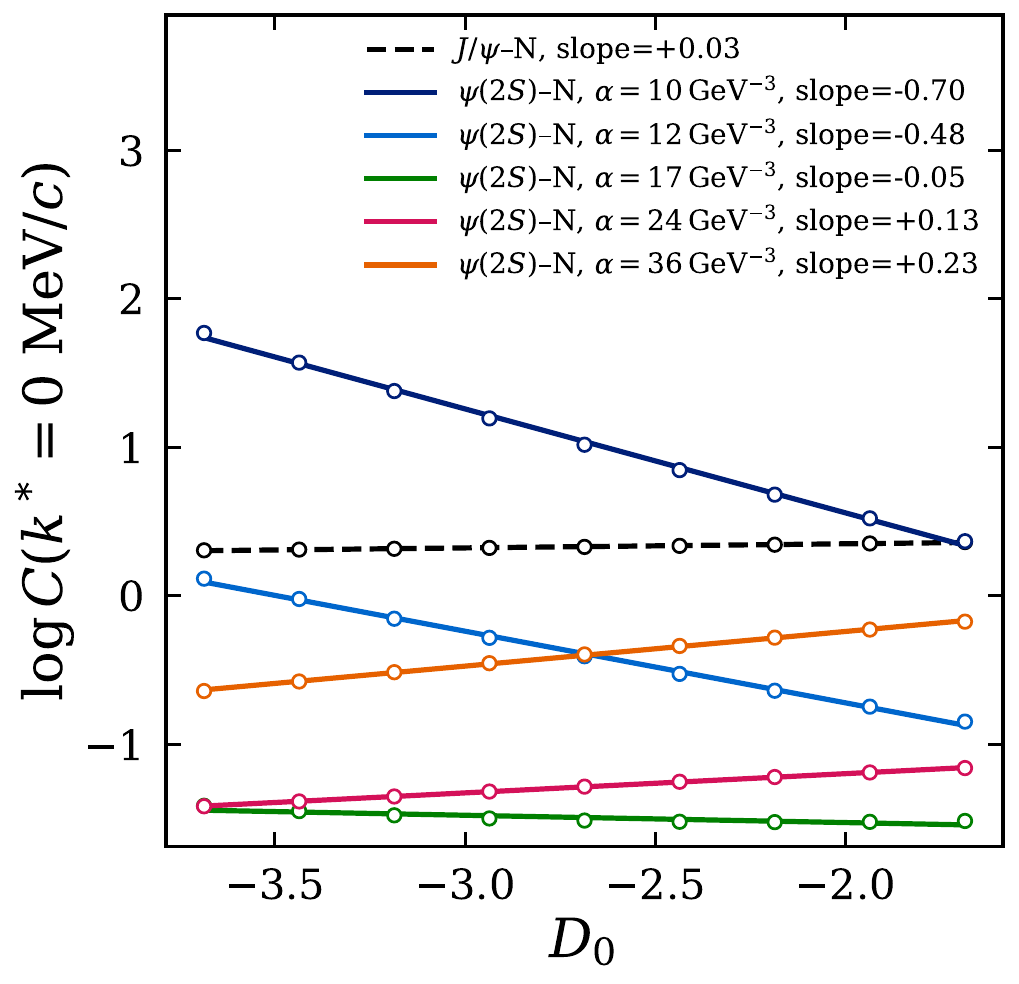}%
\caption{\label{fig:Dterm_Sensitivity}
Response of the charmonium-nucleon correlation function to the value of the $D$-term in the near-threshold region ($k^*=0$). The vertical axis is shown on a logarithmic scale.
The data points are fitted with straight lines, and the slope reflects the sensitivity to the $D$-term.
}
\end{center}
\end{figure}

In contrast to the $J/\psi$-$N$ system, lattice-QCD results for the corresponding $\psi(2S)$-$N$ potential are not currently available. Therefore, in the numerical evaluation, we vary $\alpha_{2S}$ in the range $\alpha_{2S}=10\text{--}36\,{\rm GeV^{-3}}$~\cite{PESKIN1979365, PhysRevD.94.054024, PhysRevD.98.034030}, while the other parameters, $g^2/g_s^2$ and $\nu$, are fixed to the same values as those used in the $J/\psi$-$N$ analysis. Furthermore, when evaluating the smeared potential, we use $\sigma_{\psi(2S)}=0.5\,{\rm fm}$.
With these setups, we calculate the $\psi(2S)$-$N$ correlation function, as shown in Fig.~\ref{fig:psi2SCF_Dterm}. Here, we also vary $D_0$ within the range $D_0=-2.69\pm1$.
For $\alpha_{2S}=10\,{\rm GeV^{-3}}$, the correlation function is larger than $C=1$. However, as $\alpha_{2S}$ increases, the correlation function decreases and becomes smaller than $C=1$ around $\alpha_{2S}=12\,{\rm GeV^{-3}}$. 
This behavior shows that the overall structure of the $\psi(2S)$-$N$ correlation function is mainly governed by the value of $\alpha_{2S}$. Nevertheless, the variation of $D_0$ also produces a visible change in the correlation function, unlike in the $J/\psi$-$N$ case.
In particular, this sensitivity to $D_0$ is pronounced in the small-$k^*$ region.
To see this point more clearly, we also show the same result on a logarithmic scale. 
In addition, to quantitatively evaluate the $D$-term dependence, the correlation functions at $k^* \approx 0$ are plotted against the $D$-term value in Fig.~\ref{fig:Dterm_Sensitivity}.
These log-scale plots further confirm that the dependence on $D_0$ becomes more visible in the $\psi(2S)$-$N$ correlation function, although its visibility varies with $\alpha_{2S}$.

\section{
\label{summary}
Conclusions}

In this paper, we have investigated the connection between the nucleon GFFs and the charmonium-nucleon correlation function in femtoscopy.
To describe the charmonium-nucleon system, we have employed an effective potential based on the QCD multipole expansion. In this framework, the potential is expressed in terms of the nucleon energy and pressure distributions, which are determined by the nucleon GFFs. By comparing this potential with the $J/\psi$-$N$ potential estimated from the HAL QCD data, we have examined the applicability of this effective-potential approach.
Using this potential, we have evaluated the $J/\psi$-$N$ correlation function. Since the nucleon $D$-form factor is still not well constrained, we have studied the sensitivity of the correlation function to the $D$-term. We have found that the $J/\psi$-$N$ correlation function shows only limited sensitivity to the nucleon $D$-term within the present approach. In contrast, our results indicate the possibility that
the $\psi(2S)$-$N$ correlation function is more sensitive to the nucleon $D$-term, although the relevant model parameters are not yet well determined in this case.

Recent progress in femtoscopic measurements of two-particle correlation functions, led in particular by the LHC-ALICE experiment, has provided valuable information on hadron-hadron interactions in various systems. In this context, our findings indicate that quarkonium-nucleon femtoscopy may offer a complementary way to study the nucleon GFFs. Although the present analysis does not aim at a model-independent extraction of the GFFs, it suggests that two-particle correlation functions can be sensitive to the internal mechanical structure encoded in the nucleon $D$-form factor. 

In contrast to conventional approaches based on deeply virtual exclusive processes, where GFFs are extracted through analyses of GPDs,
the present approach may provide a complementary way to constrain the GFFs from femtoscopic observables without first reconstructing GPDs. Such constraints on GFFs may also be used as additional input for GPD parametrizations, thereby helping to reduce ambiguities in the reconstruction of the three-dimensional nucleon structure  encoded in GPDs.

It is worth noting that near-threshold $J/\psi$ photoproduction may not provide a clean access to the GFFs if the observed cross sections receive sizable contributions from open-charm intermediate channels, as suggested in Ref.~\cite{Du:2020bqj}. In such a case, the connection between photoproduction observables and the soft-gluon coupling to the nucleon becomes less direct. Our approach is complementary in this respect.
In contrast to photoproduction analyses, femtoscopy does not require detailed knowledge of the initial production mechanism but instead probes the low-energy $J/\psi$-$N$ final-state interaction.
Moreover, the relevance of the soft-gluon exchange contribution to this interaction is supported by the recent analysis of $J/\psi$-$N$ scattering in Ref.~\cite{Wu:2024xwy}, which found that this contribution dominates over the open-charm coupled-channel contribution.
These observations motivate charmonium-nucleon femtoscopy as a potentially cleaner probe of the nucleon gluonic structure and GFFs.

However, to use femtoscopy as a quantitative probe of the nucleon GFFs, the model parameters entering the charmonium-nucleon potential must be better constrained. This is particularly important for the $\psi(2S)$-$N$ system, where the relevant parameters are still not well determined. This issue will be addressed in future work.

\section{data availability}
There are no publicly available research data or software supporting this manuscript. Requests for further information or data should be sent to the authors.

\begin{acknowledgments}
We are very grateful to Daniel C. Hackett, Dimitra A. Pefkou, and Phiala E. Shanahan for providing the GFFs data in Ref.~\cite{PhysRevLett.132.251904} and also very grateful to Y. Lyu, T. Doi, T. Hatsuda, and T. Sugiura for providing the $J/\psi$-$N$ potential and phase-shift data in Ref.~\cite{LYU2025139178}. We also thank Zhi-Wei Liu, Duo-Lun Ge, Jun-Xu Lu, Ming-Zhu Liu, and Li-Sheng Geng for providing the $J/\psi$-$N$ correlation function data in Ref.~\cite{3bdh-blwh}. We also thank S. Yasui and T. Gunji for insightful discussions on this work.
The work of D.F. is supported by the Japan Society for the Promotion of Science (JSPS) KAKENHI (Grants No. JP24K17054) and the COREnet project of RCNP, Osaka University.
The work of M.K. is supported by RFIS-NSFC under Grant No. W2433019.
\end{acknowledgments}


\bibliography{mybibfile}
\end{document}